# A Decentralized Approach towards Responsible AI in Social Ecosystems

**Wenjing Chu**

Futurewei Technologies, Inc.
wchu@futurewei.com

**Abstract**

For AI technology to fulfill its full promises, we must have effective means to ensure Responsible AI behavior and curtail potential irresponsible use, e.g., in areas of privacy protection, human autonomy, robustness, and prevention of biases and discrimination in automated decision making. Recent literature in the field has identified serious shortcomings of narrow technology focused and formalism-oriented research and has proposed an interdisciplinary approach that brings the social context into the scope of study.

In this paper, we take a sociotechnical approach to propose a more expansive framework of thinking about the Responsible AI challenges in both technical and social context. Effective solutions need to bridge the gap between a technical system with the social system that it will be deployed to. To this end, we propose computational human agency and regulation as main mechanisms of intervention and propose a decentralized computational infrastructure, or a set of public utilities, as the computational means to bridge this gap. A decentralized infrastructure is uniquely suited for meeting this challenge and enable technical solutions and social institutions in a mutually reinforcing dynamic to achieve Responsible AI goals. Our approach is novel in its sociotechnical co-design and its aim in tackling the structural issues that cannot be solved within the narrow confines of AI technical research. We then explore possible features of the proposed infrastructure and discuss how they may help solve example problems recently studied in the field.

## Introduction

The rise of new AI technologies promises a new era of advanced digital services that would have been impossible or impractical before. AI powered services, not only in a consumer setting (e.g., web and social media) but also in industries, public social services, and policy domains (e.g., autonomous vehicles/robotics, healthcare, housing and mortgage lending, employment, and criminal justice systems) could form the basis of our future economy and social fabric. Because of its potential impact, the AI technology's potential harms, ranging from privacy violations and social media influence operations to facial recognition in surveillance and opaque automated systems with biases, have been recognized by academia, industries, and society at large (Barocas & Selbst, 2016). Prominent research efforts studied formal and algorithmic methods of desired Responsible AI qualities including privacy (Dwork, 2008; Kairouz & McMahan, 2021) and fairness (Narayanan, 2018; Verma & Rubin, 2018; Dwork, et al., 2011; Corbett-Davies, et al., 2017). Algorithmic focused efforts alone, however, are not sufficient. If not properly deployed in a social context, technocentric solutions can suffer from common traps and fail in achieving the intended goals (Chouldechova, 2016). These shortcomings include the failure to include the crucial steps of data collection, dataset curation, and model characterization (Gebru, et al., 2018; Mitchell, et al., 2019), and more importantly, the failure to take social context into account (Chouldechova, 2016; Barabas, et al., 2020; Selbst, et al., 2019; Andrus, et al., 2021). At the same time, policy makers in various jurisdictions have recognized these risks and introduced regulations to remedy potential harms. These regulations (EU, 2018; California, 2018) will have a significant impact in AI development (EPRS, 2020) but their effectiveness is not yet evident (Machuletz & Bohme, 2020; Nouwens, et al., 2020).

To meet this challenge, we adopt a sociotechnical systems approach (Ropohl, 1999; Davis, et al., 2014) to re-frame Responsible AI problems in a new context that encompass not only the full lifecycle of AI use but also the actors and structures of a social ecosystem. This new framework gives us a robust way to discuss not just technology and people as passive users but to discuss roles and processes involving users, providers, regulators, and institutions. Based on this expansive intellectual framework, we propose a sociotechnical model for AI systems, and further propose and design a computational infrastructure, as a decentralized common

---

utility, upon which various sociotechnically effective mechanisms can be implemented to achieve Responsible AI goals.

We explore two such intervention mechanisms, agency and regulation, and discuss how the computational public infrastructure can facilitate these intervention methods and leverage social tools and dynamics to achieve Responsible AI goals. As initial steps to explore this approach, we construct a sketch of such a decentralized system building on recent advances in decentralized systems and cryptography, incrementally define a powerful set of features to solve common problems studied in the recent literature and share our thoughts and learnings about the new approach. We conclude that a decentralized approach holds great promise in advancing the balanced technical and social goals of AI with computational dynamics and policy flexibility and call for further research in this direction.

## A Sociotechnical Framework

The structural problems faced by Responsible AI do not have simple answers in the original AI technology domain (Narayanan, 2018). A sociotechnical approach (Ropohl, 1999; Davis, et al., 2014) instead seeks to optimize joint goals both on the task (functional) level and on the social level (people and their social structure). The introduction of technology into this social context creates a complex dynamic that must be understood in the combined technical and social system. This interconnectedness between a technical system and a social system is often illustrated as a diamond (Figure 1) in the classic literature (Leavitt, 1972; Bostrom & Heinen, 1977).

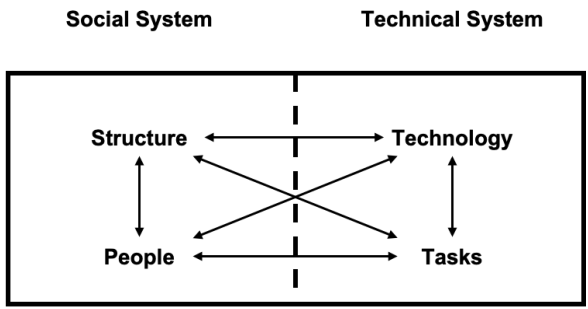

Figure 1: Sociotechnical Systems

The notion of sociotechnical systems originated from labor studies in the English coal mines after World War II (Emery & Trist, 1960). While later developments often focused on organizational studies as technology was introduced to the workplace, we believe they are still well suited as a robust framework to study the introduction of AI technology in the society. The goals of Responsible AI can therefore be understood as studying how AI technology will reshape people and their social structure, how people and their social structure will respond to AI and reshape its development and propose integrated solutions that span both the technical system and the social system for optimal outcome.

Based on this general outlook, we now discuss the following social science concepts in order to consider the requirements of a computational tool that support these social concepts.
- Human Agency

In a commercial setting, users are the subject of data collection and/or the receiver of an AI enabled service. Human agency is the empowerment of users in making self-interested decisions in a sociotechnical system. In a public policy setting, people whose data is being collected and whose lives are impacted by the AI system need to have inputs to the construction of the system and its inner workings and recourse to its automated decisions. While these two settings can have significant differences in practice, for our discussion in the high level, we will group them together.
- Regulations

We define regulations in a general sense as rules or norms constraining the operation of the technical system as well as those rules or norms that apply to the structure of people organizations. They can be algorithmic, administrative, legal, or cultural.
- Institutions

Institutions, in an abstract sense, are organizations, forums, or other digital mechanisms of people who collectively formulate regulations which make choices and compromises that prioritize certain goals over others in circumstances and make changes over time.

In sociology, institutionalization is the generalization of "value and behavior patterns," and therefore, AI technology can be seen as an example of "technical institutionalization" (Ropohl, 1999). Our central contention is that to achieve Responsible AI goals, we must design AI's technical system to foster effective institutions that formulate optimal regulations balancing task level and social level goals. A requisite condition for such effective institutions is the agency of people who are both contributors to and recipients of its impact (positive or negative) from the AI technology.

### A Sociotechnical Model for AI

Using the sociotechnical framework, we propose a simple model of common machine learning based AI systems to capture essential social actors, AI system artefacts and their relationships.

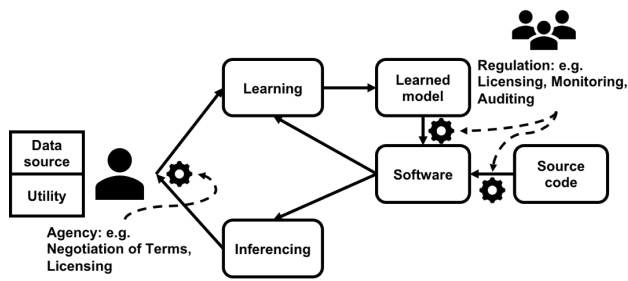

Figure 2: A Sociotechnical Model for AI

In Figure 2, the individual person, or user, is represented in the joint roles of data source and the recipient of some utility. The user makes a joint decision of contributing source data and in return receives some form of benefits, i.e., a computational utility. In current common practices, this decision process is often opaque and its ethics ambiguous as an individual user lacks practical choice and standing in negotiating the conditions of this exchange (Bohme & Kopsell, 2010; Machuletz & Bohme, 2020).

The AI system itself is modeled with a learning component and an inferencing component that interact with the user. The learning element utilizes source data from many users to algorithmically produce a trained or learned model. This model is the form where knowledge learned from the source data is codified and distributed. Typically, this model is utilized in the overall software system by combining a traditional source code program and the learned model. The combined software program is then deployed to an AI application. Similarly, learned models can also be used in a new revision of the learning algorithm itself, e.g., in a reinforcement learning setting or other forms of iterative or meta-learning algorithms.

We then propose two basic types of interventions to regulate the dynamics of the AI system with the objective to move the system towards more responsible behavior.

## Computational Agency: Empowerment

The first mechanism is to regulate the exchange between a user and an AI system. In commercial practices, this relationship is often in the forms of a Terms of Services (ToS) or End User License Agreement (EULA) for which users lack practical choices or even standing in negotiating the terms (Bohme & Kopsell, 2010; Machuletz & Bohme, 2020; Kim, 2013; Rakova & Kahn, 2020). In public service settings, the system's development is opaque, and its use is imposed upon by policy decisions where individuals of disadvantaged groups often have little input to its formulation or recourse to address its problems (Barocas & Selbst, 2016; Peacock, 2014; Chouldechova, 2016).

We call the mechanism Computational Agency because it is an empowerment mechanism in favor of the end users to own and exercise practical and effective control of their source data, and to exercise choice in the service agreement. Therefore, we symbolically put the human figure on the front side of the combined data source and utility box in Figure 2.

To make this empowerment effective, we contend that the end user must have a recognized identity to exercise such rights in the digital domain and the AI system must offer convenient enough user interface for people to exercise their rights. In social sciences, the close relationship between identity and agency is well studied (Holland, et al., 1998). In the computing domain, the Self-Sovereign Identity community (Allen & Applecline, 2017; Preukschat & Reed, 2020; Muhle, et al., 2018) offers strong arguments for universal digital identities for digital services. In the regulatory domain, the EU initiative known as eIDAS (EU, 2022) is an example of efforts now ongoing in many regions and nation states to support digital identity for their citizens.

## Computational Regulation: Rules and Norms

The second mechanism is to enforce restrictions on the behavior of the AI system. The term "regulation" is used in abstract sense here. These regulations can be imposed as public policies, ethical norms, or cultures in communities. Such regulations can put constraints on the characteristics of the datasets, a trained model's behavior regarding permissible biases, or data transparency and auditability requirements. Recent studies have proposed many accountability and auditability mechanisms and demonstrated their effectiveness (Gebru, et al., 2018; Mitchell, et al., 2019; Brundage, et al., 2020). Similarly, another aspect is to apply constraints on software code and behavior by verification mechanisms in common software distribution registries. Recourse (Ustun, et al., 2019; Joshi, et al., 2019) is another example where regulation can enforce its use in an AI-aided decision process.

In Figure 2, these regulations can be enforced most efficiently along the lines where components interact.

## A Decentralized Infrastructure

We now turn to the need to implement a practical common infrastructure and why it should be a decentralized system. While a full-scale discussion of relevant concepts and methods is out of scope for this paper, we outline a brief reasoning for our approach and offer some rationale for the proposed methods.

The first principle to consider is that of autonomy or agency from a humanistic standpoint. The challenge is how such agency can be best materialized in an AI system (in fact, any social system, digital or otherwise). The central argument is that any individual's ability to exercise meaningful equal rights in a system must start with exercising control over their own identity. For example, for a person subject to discriminatory treatment by a system to exercise the right of recourse (e.g., filing a complaint), they would first need an account, or identity, in another system independent from the very system the complaint is about. Similarly, if an online user wishes to negotiate the Terms of Service (ToS) with a provider, the basis of that negotiation must be another neutral system not subject to the very terms they are negotiating in a Catch-22.

Decentralized financial systems such as Bitcoin (Nakamoto, 2008) and Ethereum blockchain (Ethereum, 2021) make similar arguments for decentralization. However, there are significant differences. A decentralized identity is designed to exercise rights in digital systems; therefore, it is primarily concerned with neutrality among the parties, in addition to authenticity and integrity.

Such neutrality can be realized by a blockchain with the appropriate trust or governance framework, or by other forms of decentralized systems, or by systems operated by familiar social institutions that have earned such trust such as various democratic, legal, and civil institutions. In all these cases, it is a combination of a technical system and a form of governance that give it the right properties. There may be many instances of such systems that are interoperable through standardization. This is another dimension of being decentralized that ensure universality.

Such identity supporting systems must be a public infrastructure in a sense that there should be no barrier, technical or social, for individuals to create, manage, port, and remove identity and identity specific information. Neutrality requires decentralization.

Another challenge to exercise rights in a technical system is to construct an algorithmic base to efficiently establish trust, reach agreements, and verify results without a centralized authority. Many recent advances have been made in this regard for decentralized systems. Verifiable credentials (W3C-VC, 2020) allow claims or information to be asserted by the authoritative sources and be efficiently verified without an intermediary that can collect or correlate private data. Smart and Ricardian contracts (Szabo, 1994; Grigg, 2015) allow agreements to be executed as code and make its records auditable.

The right to privacy is more critical in AI systems and therefore the infrastructure we propose to exercise control over AI systems must have strong privacy support. We may consider privacy in terms of controlling collection, disclosure, storage and correlation or other inference methods in general. New signature schemes such as CL Signature and BBS+ offer efficient means to selective disclosure, Zero Knowledge Proof (ZKP) and correlation prevention (Camenisch & Lysyanskaya, 2001; Boneh, Boyen, & Shacham, 2004; Camenisch, Drijvers, & Lehmann, 2016). More advances are being made in the technologies of Homomorphic Encryption, Secure Multi-Party Computing and Secure Enclave (Cammarota, et al., 2020) that make secure and confidential computing more practical.

In addition, the public infrastructure also requires scale and robustness of a decentralized system similar to the foundation of the Internet. These and other computational mechanisms are crucial because inefficient implementations would be disadvantaged and result in the familiar ineffective rules regardless of what the text or intent of the regulation is (Utz, et al., 2019; Nouwens, et al. 2020; Machuletz & Bohme, 2020). We emphasize this point by stating that it takes a program to regulate a program.

In summary, the rationale for a decentralized system is multifold.
- It is uniquely suited to address governance issues which are at the core of sociotechnical challenges.
- It offers a practical solution to support human agency.
- It can support a set of required features for solving Responsible AI problems.
- It has the scalability and reliability needed as a common utility.

## Exploring Features and Applications for Responsible AI

To explore the strength of the proposed approach, we outline a decentralized computational infrastructure to realize the objectives we set out, namely enabling meaningful agency and implementing regulations. And we discuss how the system's features can be used to solve real-world problems that have been studied in the field. Because of the interdisciplinary nature of this work, we decide to use informal examples to discuss the system's features. While we have used less rigorous definitions of some concepts in favor of simplicity, references are provided for further research. It results in a sketch of preliminary ideas. Many aspects of the relevant technologies are also rapidly evolving and will require validation, experimentation, and revision. Nevertheless, we feel the broad strength of our approach remains valid and the essential design ideas are useful for further research in this field or practical system designs.

For the remainder of this section, we first introduce decentralized identifiers and verifiable credentials based on decentralized systems. We then sketch methods that can establish human-centric identities, can enable proof and verification with privacy, can reach binding agreements, and can enforce agreements. We speculate on market-based economic incentives and discuss various forms of governance that may be familiar in the physical world. This familiarity is an important characteristic because it helps to create and integrate reliable and durable trust models and social institutions into the AI systems.

## Decentralized Identifier

Our first objective is to create an identity for exercising agency. Decentralized identifiers (Figure 3) are a new type of globally unique identifier but avoids central administration or tracking through a decentralized system, e.g., a blockchain. While there are numerous variations, the DID Working Group in W3C (W3C-DID, 2020) is working towards standardization. W3C defines DIDs to be URIs conformant to IETF RFC 3986 (Berners-Lee, Fielding, & Masinter, 2005). In addition to being globally unique, DIDs are universally resolvable to a document which can provide basis for other properties such as access, authentication, relationship and so on. Persons or organizations may exercise control through cryptographic signature algorithms, while digital assets may use passive DIDs with an active DID as its controller.

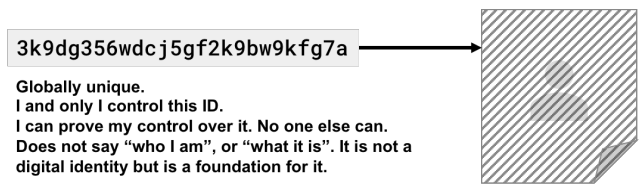

Figure 3: Decentralized Identifier

DIDs are called decentralized because the IDs can be generated and controlled (proving they have WRITE control) without relying on a centralized entity or a so-called trusted authority. Each individual can have as many DIDs as they need to reflect all the personas that they adopt in specific use cases. Through the various types of DIDs one individual may own, these systems can protect against correlation-based privacy attacks. This is a key differentiation for DIDs in contrast to other universal IDs.

Decentralized identifiers are not identities yet, but rather a root digital key that one can use to establish whatever identity or identities are needed to function and exercise rights in a digital domain. We will describe the establishment of identities later in the section.

In practical implementations, DIDs are often realized by hashing algorithms anchored in decentralized blockchains as a trust registry (Hyperledger Indy, 2020). However, other cryptographic mechanisms can also be used, e.g., KERI (Smith, 2020). Organizations can also implement DIDs through more conventional data structures combined with proper social governance mechanisms as long as they meet the trust requirements in the social context they are designed for. Such trust governance mechanisms can be achieved through social institutions. Different institutions may offer different forms of DIDs for various purposes and standards can help make them interoperable (W3C-DID, 2020).

## Verifiable Credentials

Trust is another essential ingredient for agency. Trusted information is the foundation to command & control, accountability, auditing, or reaching any basic agreement.

In addition to DIDs, decentralized systems enable issuance and verification of Verifiable Credentials (VC) (W3C-VC, 2020) and facilitate a global exchange of trustworthy information.

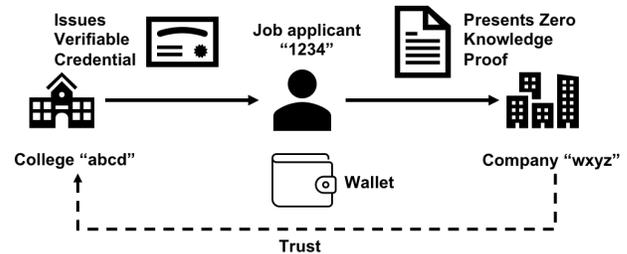

Figure 4: Verifiable Credentials

It can be best illustrated with an example. In Figure 4, we have three parties with their respective DIDs: a college with DID "abcd", a graduate of this college and a job applicant, with DID "1234", and finally a hiring company with DID "wxyz". Let us suppose the Company (DID "wxyz") may use an AI powered system to help screen candidates.

To complete a digital job application, the applicant requests a digital diploma from the College which issues a Verifiable Credential based on its private but authoritative educational records. Once received and securely stored in a digital wallet, the credential can be used to present a proof to the hiring Company. This proof is cryptographically assured based on message exchanges between the applicant and the Company without involving the credential's issuer (the College). This exchange confers the trust that the Company has with the College to the applicant even though they do not have a prior trust relationship with each other. This transitive trust relationship is fundamental in the efficient functioning of the proposed decentralized infrastructure.

The resulting system is fundamentally different from a centralized database. In our example, the applicant is the data owner and holder, who stores various credentials from many issuers to the digital wallet in their possession. Disclosure of data to the Company is fully controlled by the applicant. There is no centralized data collection about this exchange. The hiring Company only receives information relevant to the job application.

With new signature algorithms, e.g., CL (Camenisch & Lysyanskaya, 2001) and BBS+ (Boneh, Boyen, & Shacham,

2004; Camenisch, Drijvers, & Lehmann, 2016) and appropriate proof protocols, the VCs can also support selective disclosure and Zero Knowledge Proof (ZKP) to further minimize data disclosure or correlation.

**Establishing Human-Centric Identities**

A decentralized identifier is not an identity. This should be obvious (one's identity cannot be a random string of numbers and letters) but this distinction is often lost. With verifiable credentials, individuals have a mechanism to create digital identities that they choose to create to facilitate digital services and commerce. An identity consists of a set of proofs (potentially ZKPs) that are constructed from the received credentials. We emphasize that the subject and the controller of these identities are the individual (or their delegated representatives).

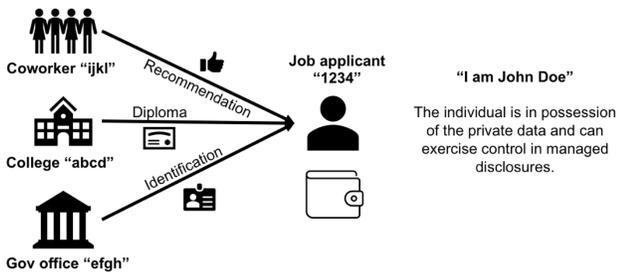

Figure 5: Human-Centric Identities

Let us continue the job applicant's example (Figure 5). In addition to the college diploma, they may request and receive a digital ID from a government office, e.g., a driver license in the U.S. which asserts their name, address, birthday, a facial photo, and some physical characteristics for identification. They may also request a letter from their previous employer for employment history and recommendations from their previous coworkers and managers, including social media recommendations such as those found in LinkedIn. These would be unsurprising credentials for a job applicant identity.

They may choose vastly different identities in different social contexts however, including being anonymous. In some digital service contexts, they may choose to construct an identity without personally identifiable information (PII) but DIDs and VCs can still assure authenticity, i.e., asserting that here is a legal person permitted to obtain this service and sign agreements or conduct transactions. In other contexts, such as the job application example above, or for obtaining government services or banking services, personal identification may be required by law or by convention. Each person can construct as many such identities as they need to obtain digital services.

**Proof and Verification**

With autonomous DIDs, and VCs issued to individuals or organizations, proof and verification can be automated and standardized. This is a key objective: to enable the easy exchange of verifiable information. This will in turn enable agreements and other forms of control.

As illustrated in Figure 6, our job applicant can present a proof using the credentials they hold in the digital wallet about their qualifications but withhold sensitive information or protect such information through ZKP to prevent biases in the applicant filtering AI system.

In a separate context, they order a drink from a Bar (DID "mnop") with a proof that they are over the legal age without disclosing other PII in their digital driver's license such as birthday and address.

Note that the verifiers, the Company, or the Bar, do not contact the original credential issuers for verification (Figure 6) reducing the risks of "phoning home".

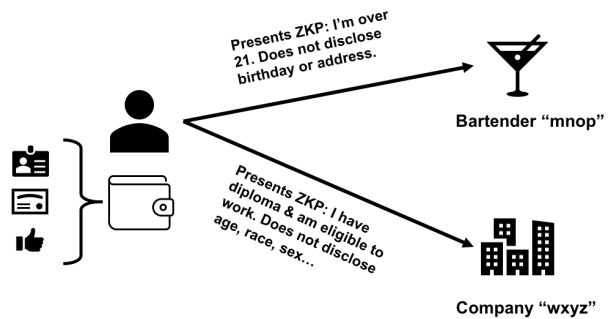

Figure 6: Privacy Preserving Proof and Verification

**Negotiating Agreements**

So far, we have outlined a set of important features provided by the decentralized computational infrastructure including autonomous DIDs and VCs and support the scalable exchange of trustworthy information, i.e., a trust layer. Now, we can discuss how parties reach agreements using this infrastructure. This will allow us to show a solution to the first problem in AI-powered systems: negotiating Term of Service.

Service exchange can be construed as a part of an agreement between parties. The previous examples, however, assume a pre-agreed protocol. This protocol can be fully digital, and standardized by law, standard bodies, or industry or community forums, and codified in software. In this section, we explore a dynamic protocol by which parties negotiate an agreement.

The basic protocol is shown in Figure 7. In this example, the service provider has been changed to an email service provider (DID "qrst"). We propose the negotiation proceed in three phases, (1) mutual identification, (2) negotiation of terms, and (3) signing.

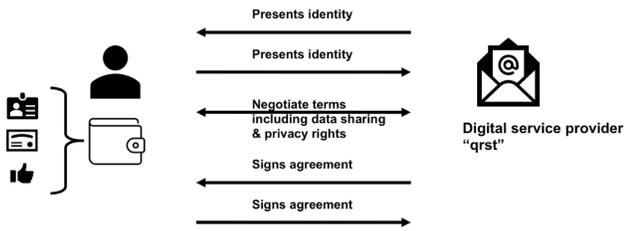

Figure 7: Negotiating Agreements

Mutual identification is straight-forward with DID and VC enabled identities. The negotiation phase consists of proposals and counterproposals between the parties to find an optimal structure. The clauses can be supported by machine readable terms (IEEE P7012, 2020) and programmable for well-known services. We propose that this structure be based on a Ricardian smart contract (Szabo, 1994; Grigg, 2000) that executes itself and is human readable and binding (Grigg, 2015; Rothrie, 2018).

Legally binding (or other forms of binding) agreements require that the digital identity and signing infrastructure are legally recognized. In recent years, some jurisdictions and institutions have been moving towards such a digital ID system (GLEIF, 2020). We argue that a decentralized identity service for all is the right approach that avoids over-centralization of power and protects individual autonomy.

How will this work for ToS negotiations? In many consumer settings, simple yet powerful methods can be (1) choice, where a user chooses one of multiple alternatives, and (2) option, where either side can propose optional add-on clauses. This will have enormous impact in the current practice of pervasive wrap contracts (Kim, 2013) that will only get more intractable with AI. With identity portability properties of DIDs, the choice and option instruments will encourage standardization, foster competitions, and be a powerful force in rebalancing a collaborative relationship between a user and an AI system to protect privacy, share service-enhancing data, and reach more optimal outcomes.

There are also more sophisticated ways of digitally negotiating Terms of Service. With structured machine-readable contracts and smart contracts, the agreements can be more nuanced taking more personal choices, and markets can be formed where data can be traded for value. In a policy setting to ensure fairness, the immutable records these smart contracts generate can aid transparency, auditing, and recourse, offline or online. Research in human-in-the-loop in AI is another important area of future studies.

**Auditing**

Auditability provides transparency so that actors in a sociotechnical system have information they need for protecting their interests and institute a reward (credit) and punishment (enforcement) mechanism. Many in research (Brundage, et al., 2020) or in policy perspective (Ada Loveless Institute, 2020) suggest auditing as a tool for Responsible AI. In social sphere, auditing supports transparency and accountability which are important in their own right for the legitimacy of a system.

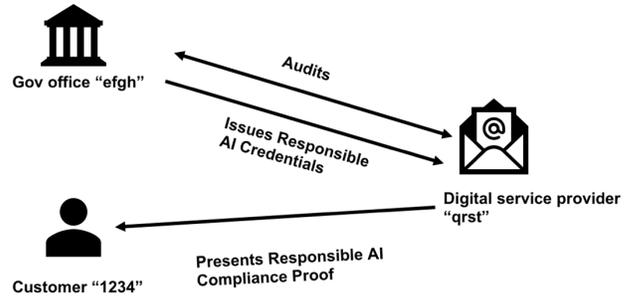

Figure 8: Auditing

The basic pattern is shown in Figure 8. The governance authority of a particular regulation conducts an audit in accordance with the said regulation. If a service provider passes the audit, a verifiable credential to that fact is issued by the authority. Then, in the service exchange setting, the service provider could offer proof of such compliance as an incentive to the customer, or the customer may request such a proof as a negotiating condition.

A verifiable log (Eijdenberg, Laurie, & Cutter, 2015) or various immutable and irrefutable data structures can also be readily implemented in scale using the same decentralized infrastructure that supports DID and VC and offer strong auditability (Brundage, et al., 2020). All Verifiable Credentials are verifiable data that can be presented with strong proofs. With any form of auditing, the verifiability of a data allows much stronger trust in the audit that does not require complex third-party arrangements (e.g., an auditor or administrative or judicial inquiry). It reduces time and cost for disagreement resolution.

Transparency can be enhanced with the decentralized infrastructure's assist. Disclosure of internal data for auditing is often hampered by the need to protect proprietary information as well as privacy for those involved. With strong anonymization features built-in and efficient secure multi-party computing (S-MPC), a consumer or justice advocacy group, e.g., can conduct rigorous verifiable audits without directly accessing the underlying data. Other types of auditing methods such as sock puppet (Asplund, et al., 2020) can be vastly scaled with authorized sock puppet DIDs.

## Portability

Autonomous identities can be used as an address for communication services such as email, messaging, or social media. Emails would be addressed to "John Doe" (DID 1234) rather than john.doe1234@qrst.com (Figure 9). With individually owned portable addresses, service providers will be less likely to become self-interested monopolies that users cannot practically leave (Gans, 2018; Windley, 2005).

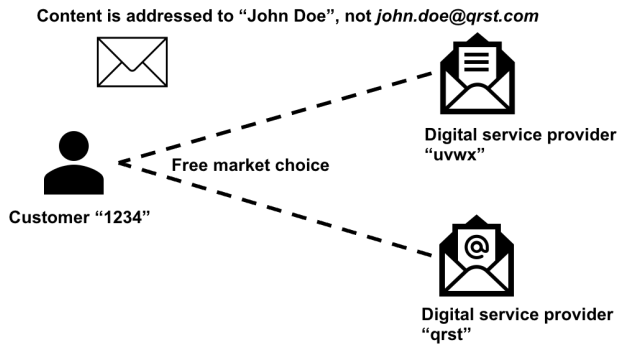

Figure 9: Portability

With portability, they can more practically exercise free market choices that therefore put pressure on the provider to practice Responsible AI out of its own best interest. In such a system, users maintain full control of their own email addresses which are completely separate from the services offered by the mail providers. Customers can readily "vote with their feet" if they are dissatisfied with aspects of the service including the handling of private information and the availability of AI powered capabilities. The basic trade of a user's data for enhanced services can still function, but the balance of power now favors a fairer trade. Pooling data from a large number of users will improve AI performance and gain competitive advantage, therefore the economy of scale and competitive incentive continue to function in such a marketplace.

Combining portability with negotiation of Terms of Service, we may have a system that is fairer, more competitive, and advantageous to the long-term development of AI technology. We argue that if we are to meet the Responsible AI challenge, we must not leave these important factors out of our research agenda.

## Institutions

Finally, we discuss the governance structure of the decentralized system we described in this paper and highlight how it can help the establishment of future digital institutions. Institutions, in a sociotechnical sense, are social pacts or norms that people form to integrate technology into a social environment. They are therefore crucial in shaping the trajectory of AI development. Agency allows people to form such institutions to advance their common interests. The computational infrastructure we propose merely helps to make these institutions efficient and scalable in an AI ecosystem.

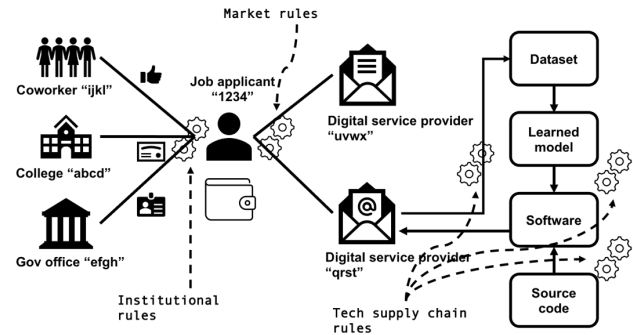

Figure 10: Institutions

Let us discuss a few examples (Figure 10) of such social governance pacts in order for the proposed technical system to work as intended towards Responsible AI.

The first of these is the rules governing the issuance of credentials. The college in our original job applicant example derives its authority to issue diplomas from its legal charter, accreditation, and continuous responsible exercise of such authority. Similar reasoning also applies to individual recommendation letters. A government office may derive such authority through political means. We believe that social institutions such as these will continue to play their roles in digital services, but more importantly, new types or modified forms of institutions will emerge to meet the new demands that are specific to AI-powered services.

The second example is the regulation of the proper functioning of markets. The recent rapid developments in digital currency, assets, and markets opened a new way to study market dynamics in a digital system.

The third example is in the regulation of technology businesses, e.g., anti-trust (OECD, 2017; Ezrachi & Stucke, 2016). In these areas, regulators may impose a structure and enforce rules between the boundaries. As shown in Figure 10, we may regulate

- data collection,
- learned model's biases, and
- the supply chain of source code and other components.

In each of these examples, a decentralized approach strengthens the AI system's accountability and incentivizes Responsible AI with flexibility for policy choices.

## Conclusions and Related Work

In this paper, we explore a strong sociotechnical approach to tackle the central challenges we face in Responsible AI.

Our novel approach differs from algorithm-centric research seeking optimal performance of an abstracted task and also differs from social-aware research where algorithms are enhanced to meet formally defined privacy or fairness constraints while optimizing the task's performance. Instead, our contribution is to propose a strong sociotechnical co-design approach that puts AI technology and the social actors who develop and use the technology in a unified framework and seek a system dynamic that can produce the desired outcome.

With that framing, we outlined a sociotechnical model to describe common AI systems. This model is unique that it brings in social actors such as users, providers, the public, and regulators into the scope of study and captures artefacts such as datasets, trained models, and software of the AI system. Guided by social science concepts, we identified two intervention mechanisms: agency and regulation, and incorporated them into the model.

To realize such a model in Internet-scale, we proposed a decentralized public utility for the purpose of regulating AI system behavior within the proposed framework. This decentralized utility is the infrastructure to materialize the sociotechnical constructs.

With that foundation, we incrementally sketched out a rich set of features for the system. These features include decentralized identifiers, verifiable credentials, human-centric identities, agreements, auditing, portability enabled market mechanisms, and digital governance institutions. These features are powerful tools, and they are unique in how they exploit the *system dynamics* in a sociotechnical system between technology and ecosystem and among the system's social actors. In sociotechnical co-designing, we seek reinforcing dynamics and policy flexibility to achieve optimal equilibrium. We explored how these features can address challenging problems related to privacy, user autonomy, transparency, accountability, fairness, and recourse. While these feature designs are preliminary, we offered insights and demonstrated a novel sociotechnical co-design approach towards solving Responsible AI problems. These features are promising areas for further experimentation and studies.

## Related Work

A rich set of recent studies have advocated a sociotechnical approach in AI research and AI related policy making. Selbst, et al. (2019) identify common conception traps related to fairness AI research. Barabas, et al. (2020) characterize AI and data science as a sociotechnical process that is "inseparable from social norms, expectations and contexts of development and use." Andrus, et al. (2021) call for reevaluation of problematic technical abstractions that researchers and practitioners have assumed in AI and argue for reframing the AI fields to include human and social factors to model the full system of interest. Poechhacker & Kacianka (2021) identify that formal expression of causality as a means of AI accountability must be understood in a social context. And the classic sociotechnical studies from the 1950's at the time of the introduction of industrial-age technologies still resonate strongly today (Emery & Trist, 1960). Our work continues in this direction, and it is novel in its strong sociotechnical approach to tackle the structural problems directly rather than improving the technical systems within its domain abstractions, or impact and policy studies sorely on the social system side. We also propose decentralized systems as the ideal means uniquely suited for this purpose where social concepts like agency and regulation can be efficiently introduced in the technology institutionalization process.

Combining decentralized blockchain systems with AI has also seen significant interests in recent years. But the focus is often on data sharing while preserving ownership or complying with privacy regulations through federated learning settings (Cheng, et al., 2019; Harris & Waggoner, 2019; Kairouz & McMahan, 2021). Many others have suggested to use blockchain for diverse purposes of data provenance, data authenticity, system reliability and more (Salah, et al., 2019). These are designed mainly as enhancements to the technical AI system. Our proposed decentralized infrastructure is novel as it is designed for the social goals of empowering human users and fostering AI-age social institutions in regulating AI systems.

Providing a common computing utility to solve complex problems is not a new idea. Several decentralized systems are in operation today with the goal of providing universal identity with strong privacy features (Sovrin, 2021; GLEIF, 2020; Hyperledger Indy, 2020). The design of these systems put a great deal of thoughts into crafting a decentralized governance structure. Their work inspired us to apply what we learned to solving problems in Responsible AI. Successful blockchain based financial systems also offer opportunity to study the interplay between technical and social systems (Nakamoto, 2008; Ethereum, 2021). In other technical domain, both PKI and DNS can be thought of as such common infrastructures although neither is decentralized which caused many problems we are to address. For software developers, the ubiquitous Github service is an example of how a commercial enterprise may be incentivized to support such an infrastructure. And obviously, the Internet infrastructure itself is fully distributed and partially decentralized, built as a public infrastructure.

In writing of this paper, we recognize the enormous challenge in a complex interdisciplinary study that crosses many technical fields and social science fields. However, we believe such interdisciplinary approach is important for AI development and hope our work can spur interests and future research in this direction.

## Acknowledgements

We wish to thank all members of the Trustworthy Intelligent Computing (TIC) project and in particular Brice Dobry for his reviews and discussions and the participants of the AAAI 2020 Spring Symposium for their useful feedback. We thank the invaluable feedback to an earlier draft from the anonymous ICWSM reviewers.